\documentclass[letter]{aa} 
\usepackage{color}
\usepackage{graphicx}
\usepackage[]{natbib}
\usepackage{sidecap}
\usepackage{txfonts}
\newcommand{\JCoPh}{  {\it J. Comp. Phys.}}

\begin{document}
\title{Recurrent solar jets in active regions.}

\author{ V. Archontis$^1$
        \and
        K. Tsinganos$^2$
        \and
        C. Gontikakis$^3$}

\institute{School of Mathematics and Statistics. St. Andrews University, St. Andrews, KY16 9SS
         \and
Section of Astrophysics, Astronomy and Mechanics, Department of Physics, University of Athens, Panepistimiopolis, Zografos 157 84, Athens, Greece
         \and
Research Center for Astronomy and Applied Mathematics, Academy of Athens, 4 Soranou Efessiou Str., Athens 11527, Greece}
\date{Received 27 November 2009; Accepted 24 February 2010}

\abstract
{}
{We study the emergence of a toroidal flux tube into the solar atmosphere and its interaction with
a pre-existing field of an active region. We investigate the emission of jets as a result of repeated reconnection
events between colliding magnetic fields.}
{We perform 3D simulations by solving the time-dependent, resistive MHD equations in a highly stratified atmosphere.}
{A small active region field is constructed by the emergence of a toroidal magnetic flux tube.
A current structure is build up and reconnection sets in when new emerging flux comes into contact with the
ambient field of the active region.
The topology of the magnetic field around the current structure is drastically modified during reconnection. The modification
results in a formation of new magnetic systems that eventually collide and reconnect.
We find that reconnection jets are taking place in successive recurrent phases in directions perpendicular to each other, while in each phase they
release magnetic energy and hot plasma into the solar atmosphere. After a series of recurrent appearance of jets, the system approaches an
equilibrium where the efficiency of the reconnection is substantially reduced.
We deduce that the emergence of new magnetic flux introduces a perturbation to the active region
field, which in turn causes reconnection between neighboring magnetic fields and the release of the trapped energy in the form of jet-like emissions.
This is the first time that self-consistent recurrency of jets in active regions is shown in a three-dimensional experiment of magnetic flux emergence.}
{}

\keywords{Magnetohydrodynamics (MHD) -- Methods: numerical -- Sun: activity -- Sun: corona --
Sun: magnetic fields}

\maketitle

\section {Introduction}
\label{sec:intro}
Jet-like emissions of plasma in the solar atmosphere have been extensively observed over a range of wavelengths
(X-Ray, EUV, H$\alpha$). They usually occur in active regions and polar coronal holes. It is believed that many jets
and surges are produced directly by magnetic reconnection \citep{shibata07} when oppositely directed magnetic
field lines come into contact. Due to reconnection, the magnetic energy of the fields is converted into heat and
kinetic energy of the ejected plasma. Observations \citep{chae99} have also shown that EUV jets 
and H$\alpha$-surges occur in regions of magnetic cancellation between 
emerging and pre-existing magnetic fields of opposite polarity. Thus, the idea that the jet formation is due to 
an interaction of magnetic fields is widely supported by various measurements and numerical experiments 
\citep[e.g.][]{yoko96, arc05, fmi08}\\
In many cases, the appearance of jets is recurrent. \citet{chi08a,chi08b} have shown a recurrent jet emission in an active region.
They found that the emission was associated with magnetic flux cancellation and they suggested that the emission was coming from the
chromosphere in the process of evaporation. \citet{chae99} suggested that magnetic reconnection
driven by emerging flux would be a possible scenario for the recurrency of EUV recurrent jets in active regions.
\citet{wan02} found numerous jet-like ejections, originated from active regions
located inside or near the boundaries of nonpolar coronal holes. The jets were apparently triggered when the magnetic loop systems of the active region
reconnected with the overlying open flux.\\
\citet{murray09} studied the emergence of magnetic flux in a coronal hole via 2.5 MHD numerical simulations. They found that oscillatory reconnection
occurs between the rising field and the open ambient field of the coronal hole environment. A cyclic evolution of temperature and repeated reconnection
outflows were reported as a consequence of the oscillatory reconnection. \\
In a previous study \citep{gont09}, hereafter paper I, we showed the formation and emission of a reconnection jet, driven by the emergence of a toroidal
loop at the edge of an active region. The physical properties of the jet were in good qualitative and quantitative agreement with observations of an
active region jet. Here we study the long-term evolution of the system focusing on the characteristics of the reconnection process and the jets.
We find a persistent behavior of reconnection (similar to the 2.5D oscillatory reconnection by \citet{murray09}) between the interacting magnetic
fields and also recurrent emission of jets.
This is the first reported instance of recurrent jets in 3D, driven by reconnection that is initiated by flux emergence into pre-existing closed
loops of an active region. \\

\begin{figure*}[t]
\centering
\includegraphics[width=4.5cm,height=5.0cm]{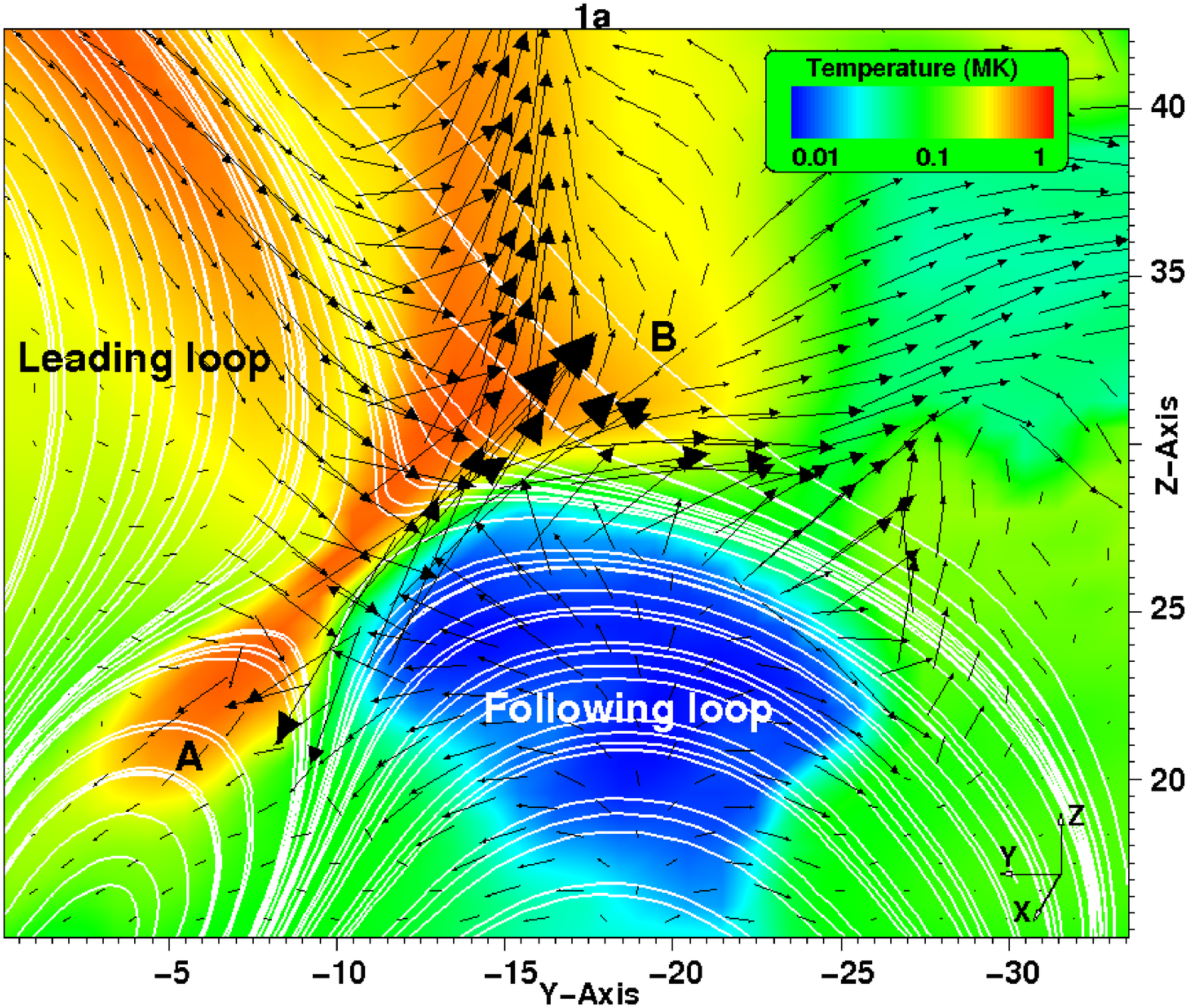}
\includegraphics[width=4.5cm,height=5.0cm]{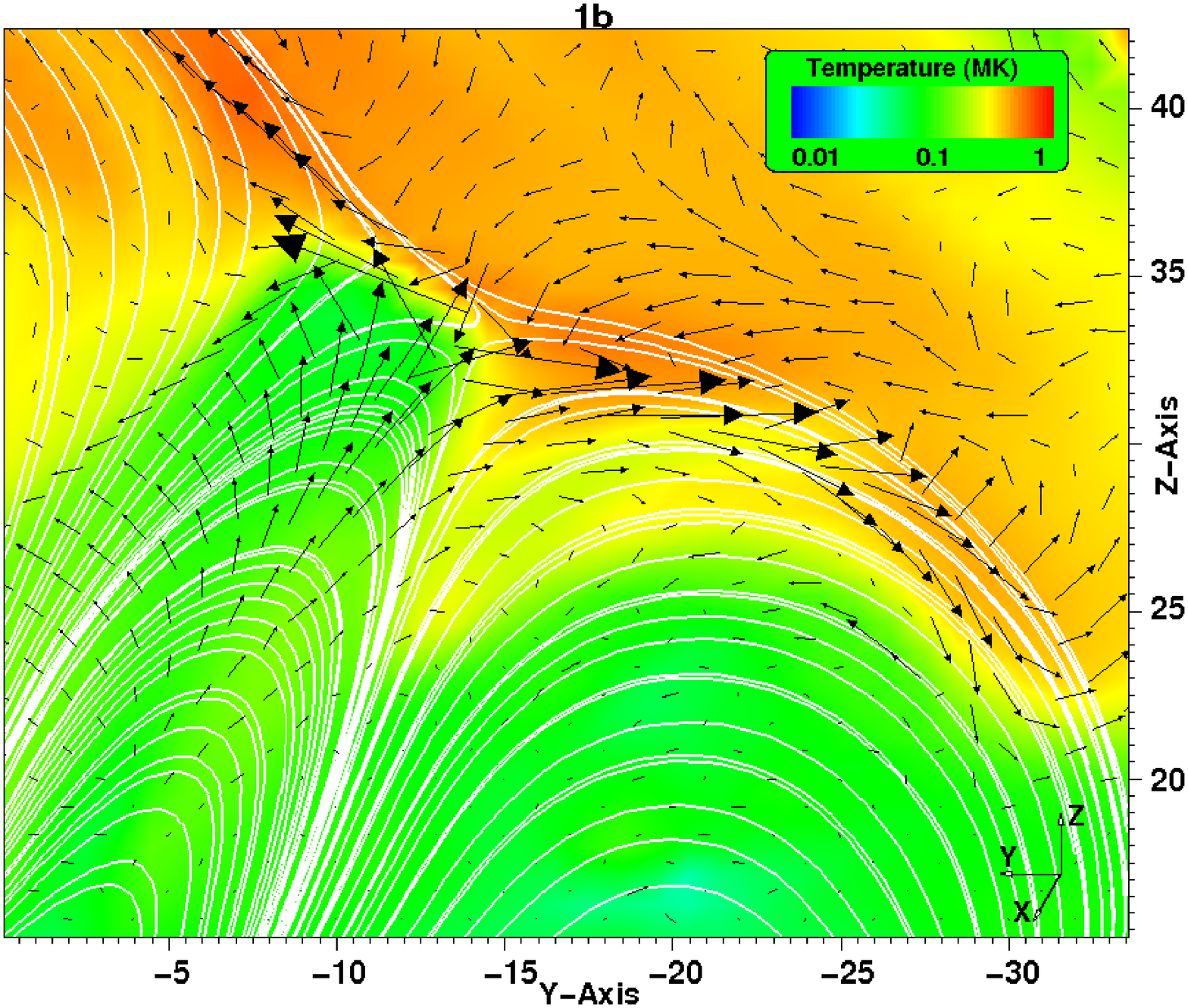}
\includegraphics[width=4.5cm,height=5.0cm]{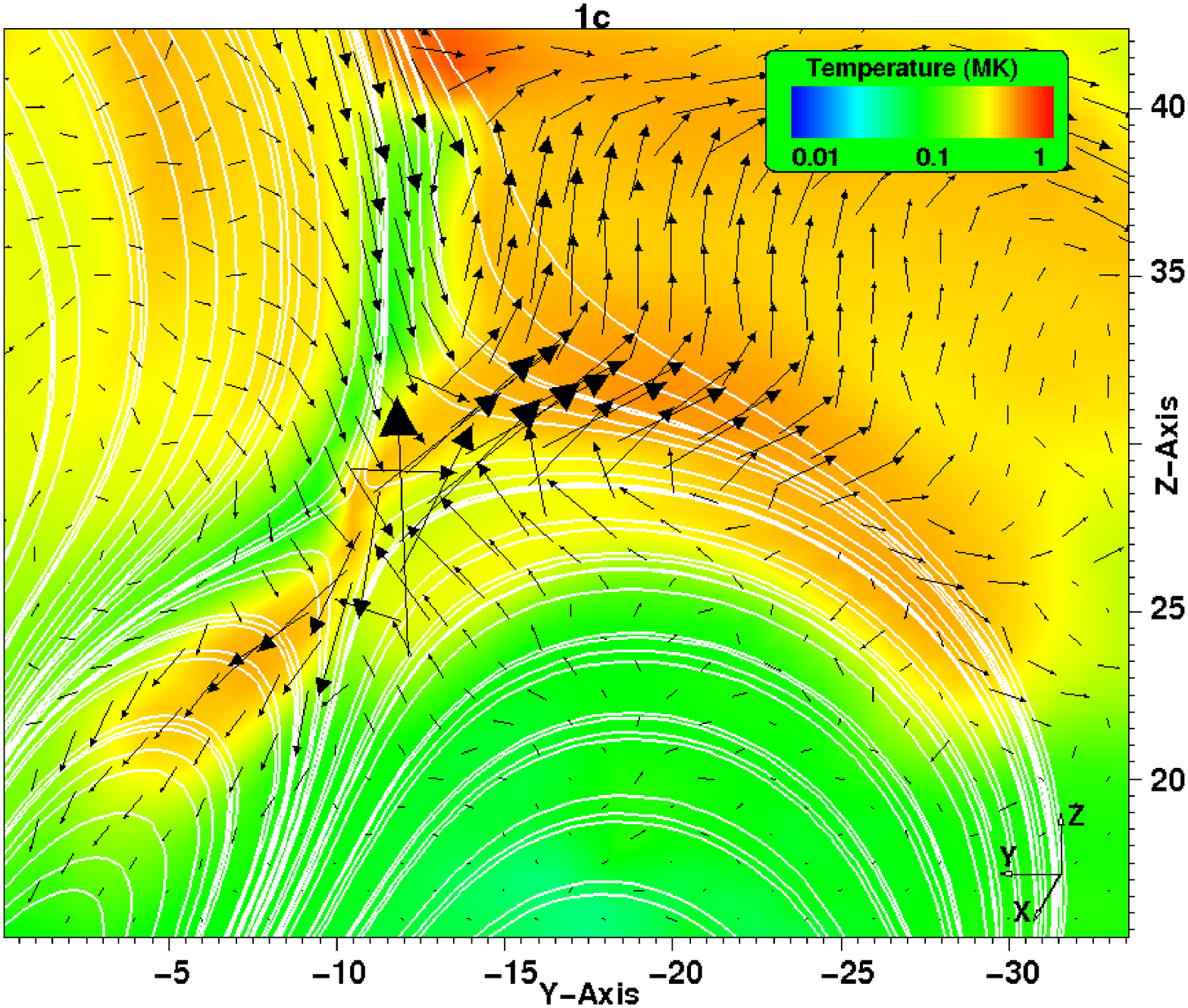}
\includegraphics[width=4.5cm,height=5.0cm]{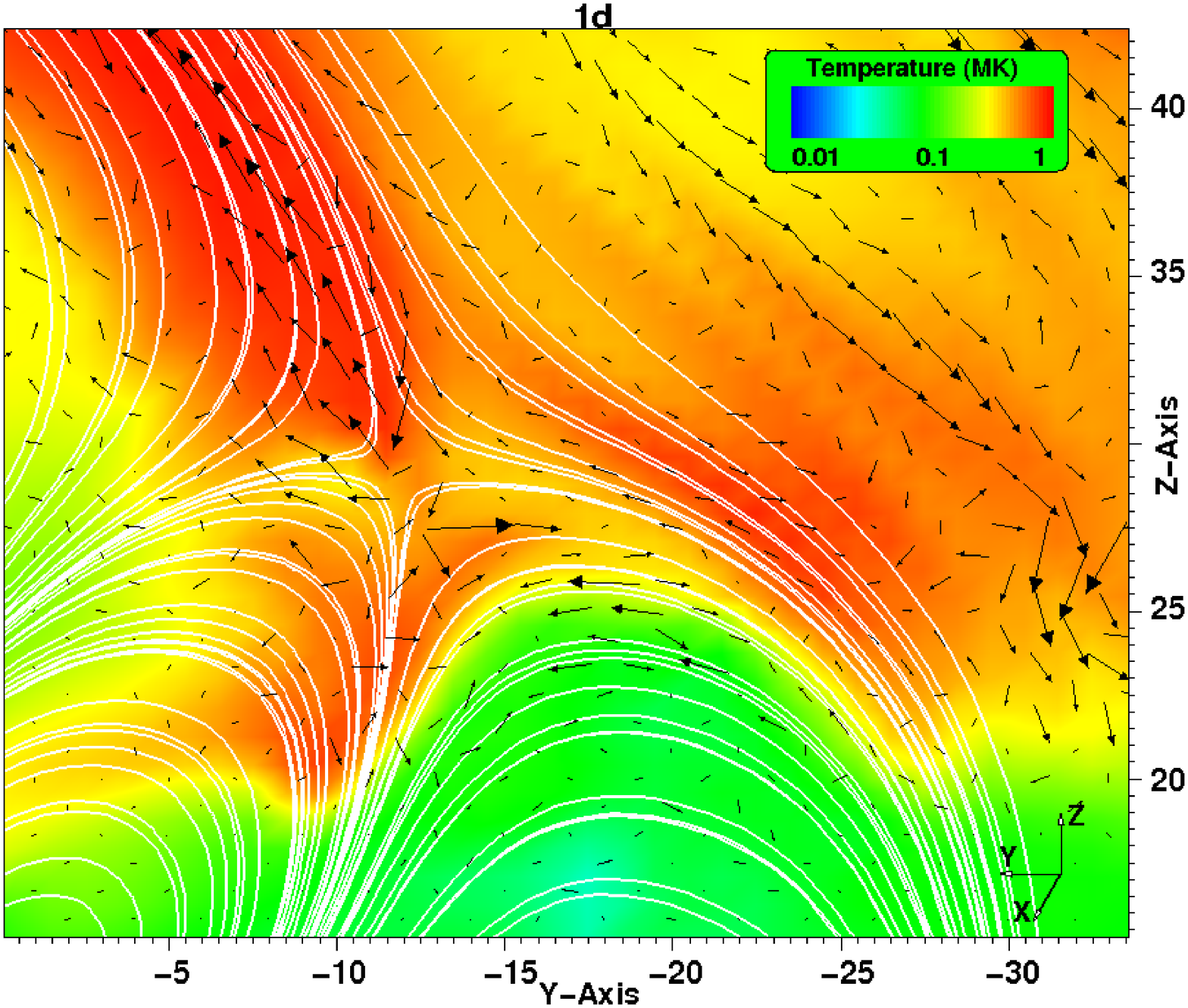}
\caption{ Recurrent emission of jets due to reconnection at $t=144$ (1a), $t=184$ (1b), $t=192$ (1c) and $t=228$ (1d) in the $x=10$ plane. 
Specifically, in panels 1a and 1c jets are emitted in the 1 and 7 o'clock directions, toward the {\it arcade} and {\it envelope} loops. 
In panels 1b and 1c jets are emitted in the 4 and 10 o'clock directions while there is inflowing plasma from the {\it arcade} and 
{\it envelope} loops. Color contours show the temperature distribution. Velocity field (arrows) and magnetic field lines 
(white lines) are overplotted onto the plane.}
\end{figure*}

\section {Model}
\label {sec:model}
The results in our experiments are obtained from a 3D magnetohydrodynamic simulation using a
Lagrangian remap scheme \citep{arber01}.
The basic setup follows the simulation in paper I.
The initial state consists of an hydrostatic atmosphere and
two toroidal magnetic flux loops.
All variables are made dimensionless by choosing photospheric values for the density,
$\rho_{\mathrm{ph}}=3\times10^{-7}\,\mathrm{g}\,\mathrm{cm}^{-3}$, pressure,
$p_{\mathrm{ph}}=1.4\times10^{5}\,\mathrm{ergs}\,\mathrm{cm}^{-3}$, and pressure scale height,
$H_{\mathrm{ph}}=170\,\mathrm{km}$, and by derived units (e.g., magnetic field strength
$B_\mathrm{ph}=1300\,\mathrm{G}$, velocity $V_\mathrm{ph}=6.8\,\mathrm{km}\,\mathrm{s}^{-1}$ and
time $t_{\mathrm{ph}}=25\,\mathrm{s}$).
As in paper I, the atmosphere includes a subsurface layer $(-25\le z<0)$,
photosphere $(0\le z<10)$, transition region $(10\le z<20)$ and corona $(20\le z\le 100)$. The (dimensionless)
size of the numerical domain in the longitudinal ($y$) and transverse ($x$) directions is $[-80,80] \times [-80,80]$.
The toroidal loops are imposed below the photosphere along the $y$-axis. The crest of the first toroidal loop
must rise $1.2 Mm$ to meet the surface ({\it leading} loop). The corresponding distance for the second loop is $2.2 Mm$ ({\it following} loop). To initiate
the emergence, the entire loops are made buoyant by setting the temperature within the tubes equal to the temperature of the
background atmosphere. The density deficit and excess pressure along the loops have been introduced in \citet{hood09} and in paper I.
The first magnetic elements of the {\it leading} tube that reach the photosphere have a field strength of $\approx 2 KG$. The {\it following}
loop arrives at the photosphere with a weaker field strength, around $1.7 KG$. The values for the twist, the
minor and the major radius of the toroidal flux loops are the same as those of paper I.

\section {Results}
\subsection {Recurrent appearance of jets}
\label{sec:recurrency}

\begin{figure*}[t]
\centering
\includegraphics[width=18cm,height=7.0cm]{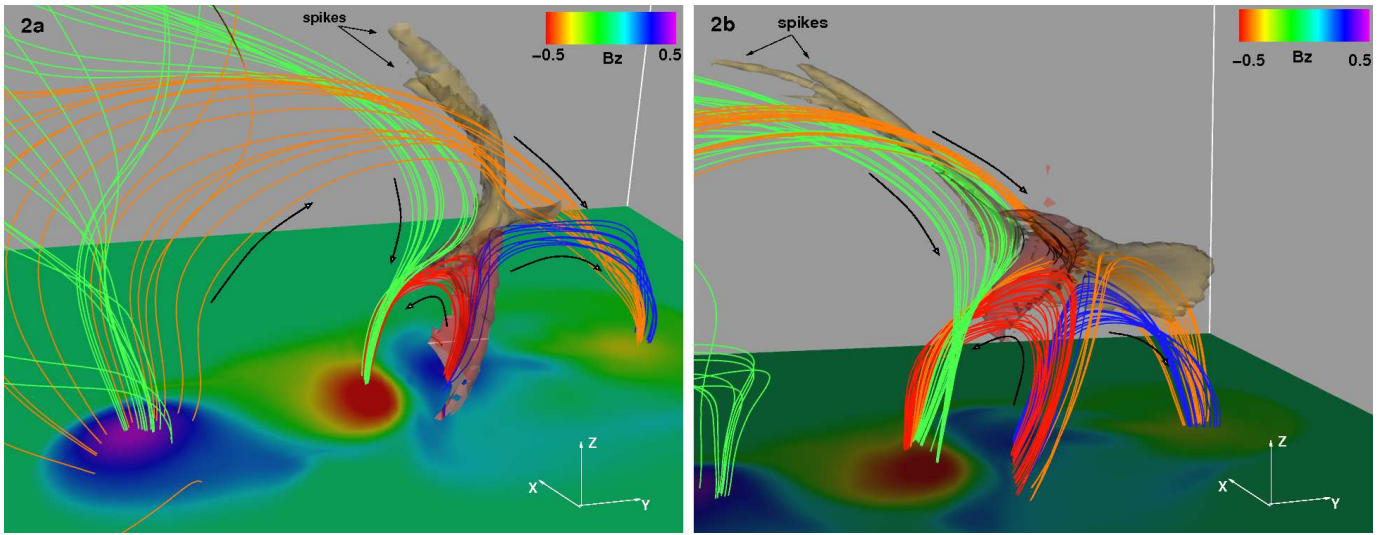}
\caption{3d visualization of the jets (velocity isosurfaces, yellowish/grey) at $t=144$ (left) and $t=184$ (right). Side views are shown for the two snapshots.
The current sheets (colored red) are visualized by calculating {\bf J/B}. The horizontal slice is a {\it magnetogram} at
$z=2$. Note that the two upward elongated jets are emitted along similar directions (oblique-left). The arrows (black color) show 
the direction of the full magnetic field vector.}
\end{figure*}

Figure 1 (panels 1a-1d) shows the emission of bi-directional flows (jets) at four different times during the evolution of the
system. The colored slice is a 2D horizontal plane ($x=10$) that shows the distribution of the temperature. Shown also is the projection of the
full velocity vector onto the plane (arrows) and the magnetic field lines (white lines). At $t=144$, the
{\it leading} magnetic loop has risen well into the corona, producing an external ambient field for the {\it following} loop to come into. When the
two loops meet, a current layer is formed at their interface ($-8<y<-12$, $24<z<28$). The fieldlines on the two sides of the interface are oppositely
directed and, thus, they reconnect along the current structure. The reconnected fieldlines form two new magnetic domains, above
and below the edges of the interface (marked with {\bf A} and {\bf B}, panel 1a). The fieldlines in domain {\bf A} form an {\it arcade}-like
structure.
The domain {\bf B} consists of an {\it envelope} field that overlies the rising toroidal loops.
The dynamical interaction between the four magnetic domains
is important for the recurrency of jets.\\
It was shown in paper I that the initial interaction between the two magnetic loops leads to the formation of a hot and
high-velocity reconnection jet. This is also shown here in panel 1a. Firstly, the lateral expansion of the {\it leading} loop and the adiabatic, rising
motion of the {\it following} loop drive inflows that bring magnetized plasma into contact. Then the magnetic
field lines that press against each other reconnect, producing outflows (jets), which are directed towards the {\it arcade} and {\it envelope} fields. 
The bidirectional flows are accelerated by the tension force of the reconnected fieldlines. This is the first episode, but not the
last, in the dynamical evolution of magnetic fields that results in jet formation. The velocity of the jets may reach values in the range
$100-200 Km/sec$. The emitted hot plasma has temperatures of a few MK during the evolution.\\
Eventually (panel 1b) the topology of the flow around the reconnection site experiences a substantial change.
The {\it arcade} field undergoes an apparent vertical expansion due to the addition of reconnected fieldlines at the top of the {\it arcade}. In this way the
crest of the {\it arcade} approaches the {\it envelope} field, forming a new current layer, which is located higher in the atmosphere ($-11<y<-15,33<z<36$)
and its cross section undergoes a rotation by $\approx 90$ degrees relative to the interface at $t=144$. This time it is the fieldlines
of the {\it arcade} and {\it envelope} fields that reconnect to produce jets. 
The reconnection jets move toward the two emerging fields, which were
previously possessing inflows. The reversal in the direction of the velocity flow leads to new reconnection events
and a recurrency of jets. During the experiment we are witnessed two more episodes of reconnection outflows, at $t\approx192$ (panel 1c) and $t\approx228$ (panel 1d). 
The change in the topology of the flow field occurs alternately: at $t=144$ and $t=192$ the inflow regions are the emerging toroidal loops, while at
the intervening time $t=184$ and $t=228$ the inflows emanate from the {\it arcade} and {\it envelope} fields.\\
The physical properties of the recurrent jets change over time. Their velocity, for example, does not appear so high in all episodes.
In the last event the bidirectional outflows do not have speeds more than $50 Km/sec$. The temperature along the jets may also drop from
a few MK during the first ejection to $\approx 500.000 K$ in the last emission. At that stage of evolution, the apparent enhancement of temperature
along the reconnection outflows is also due to the compression of neighboring magnetic fields. The initial plasma density of the jets is more than ten times the
density of the background atmospheric plasma. The density in the following jets may decrease by a factor of 2. After $t=240$, the system approaches
a stage where the occurrence of jets is drastically diminished. The recurrent jets do not have the same properties because the magnetic 
systems that come into contact have a  specific initial reservoir of magnetic flux and energy. Each time they reconnect, energy is released 
and the flux is eventually exchausted. Consequently, each reconnection event between the same magnetic flux systems is less effective than the 
previous one. As a result, the recurrent jets appear to have different physical properties (e.g. temperature).\\
Figure 2 shows the three-dimensional emission of the jets. At $t=144$ (panel 2a) the {\it leading} and {\it following} loops (green and blue fieldlines respectively)
reconnect along the current structure (transparent red isosurface), which adopts an arch-like shape.
The orange field lines that join the positive polarity of the {\it leading} loop with the negative polarity of the {\it following} loop represent the magnetic
domain of the {\it envelope} field. The {\it arcade} magnetic field is shown by the red fieldlines.
At the beginning of the emission, the jet is directed vertically above the current, but eventually it becomes collimated
along the reconnected fieldlines of the {\it envelope} field. The 3D visualization reveals that the jet adopts a double-peak structure (panel 2b).
The spikes are developed at the leading edge of the jet and are moving along parallel fieldlines that belong to the same ({\it envelope}) field.
At $t=184$ (panel 2b), the jets are emitted sideways from the rims of the current structure. They are directed on opposite sides along the ambient fieldlines.\\ 
We note that the initial emission of the recurrent jets occurs in perpendicular directions. Eventually, the upward jets move along 
the reconnected fieldlines that envelope the leading loop. As a result, the final direction of these jets is similar: they are pointing along 
the negative direction of the y-axis, in an oblique-left orientation. On the other hand, the relative orientation of the downward jets makes 
an angle of $\approx 90$ degrees. No specific observations to date have indicated this feature. We believe that the geometry of the overall 
system plays an important role in determining the final direction of the jets. If, for example, the emerging field 
reconnects with a (constant and uniform) pre-existing vertical (or oblique) field the direction of all recurrent jets will 
be vertical (or oblique). If on the other hand the pre-existing field evolves dynamically into the 3D space the direction of the 
jets depends on the relative orientation of the fieldlines of the magnetic systems at the time of their contact.
\subsection {Driving mechanism}
\label{sec:Driving mechanism}
Now we study the effect of the reconnection process on the recurrency of the jets.
Figure 3 shows the time evolution of the maximum current density {$\bf J = |{\bf \nabla} \times {\bf B}|$} at the evolved current structure between
the interacting magnetic fields. For the calculation we measured the maximum {\bf J} within the current structure.
We also plot the maximum value of the parallel electric field $E_{\parallel}$ (in the same region), which is a rough estimate
of the reconnection rate between the magnetic fields into contact.
Based on the reversal of the flow topology around the diffusion region, it is possible to distinguish four reconnection phases (RP1 to RP4).
In each phase, {\bf J} first (initial stage) reaches a maximum value and then (later stage) drops before the next flow
reversal. The duration of each
phase is between $9$ and $13$ minutes.
In RP1 and RP3 inflows bring the two emerged toroidal loops into the diffusion region. In RP2 and RP4 the
inflows to the current structure and the outflows from the diffusion region have a reversed direction. 
Figure 1 shows the emission of the reconnection outflows at a time when the value of {\bf J} is maximum in each reconnection phase.\\
Figure 3 shows that there is a good correlation between {\bf J} and $E_{\parallel}$. During the initial stage of RP1 ($130<t<144$), the {\it following}
loop is emerging and comes into contact with the pre-existing field for first time.
Thus, the magnetic stresses throughout the region of the interface (mainly on the side of the {\it following} loop)
are large, increasing the compression at the interface. As a result, the current density is enhanced, reaching a peak value
at $t=144$. $E_{\parallel}$ follows a similar evolution. The reconnection rate is minimal during the initial stage, but thereafter
it increases as the current structure builds up at the interface.
As the fieldlines diffuse in through the plasma and cancel, the
fluid is expelled out of the ends of the current layer, which eventually dissipates (later stage). Consequently, the reconnection becomes
less effective and the reconnection rate quickly drops to a low value. This is shown by the decrease of $E_{\parallel}$ in
the later stage of RP1.\\
The next time that these two systems press against each other is during RP3.
Compared to RP1, the maximum {\bf J} is smaller. The same trend in the
evolution of the current is followed during the
other two phases: {\bf J} is larger when fieldlines from the {\it arcade} and {\it envelope} fields reconnect for the first time (RP2) and smaller
during the second time (RP4). Also, it is stronger during the reconnection between the main emerging fields 
(RP1 and RP3) and weaker during RP2 and RP4.
In all phases, the reconnection rate undergoes a parallel evolution to the current density. They develop a similar
trend and reach local maxima and minima at approximately the same time.
Their behavior after RP4, might be described as convergent evolution towards an equilibrium.\\
\begin{figure}
\begin{center}
\includegraphics[width=8cm,height=6.cm]{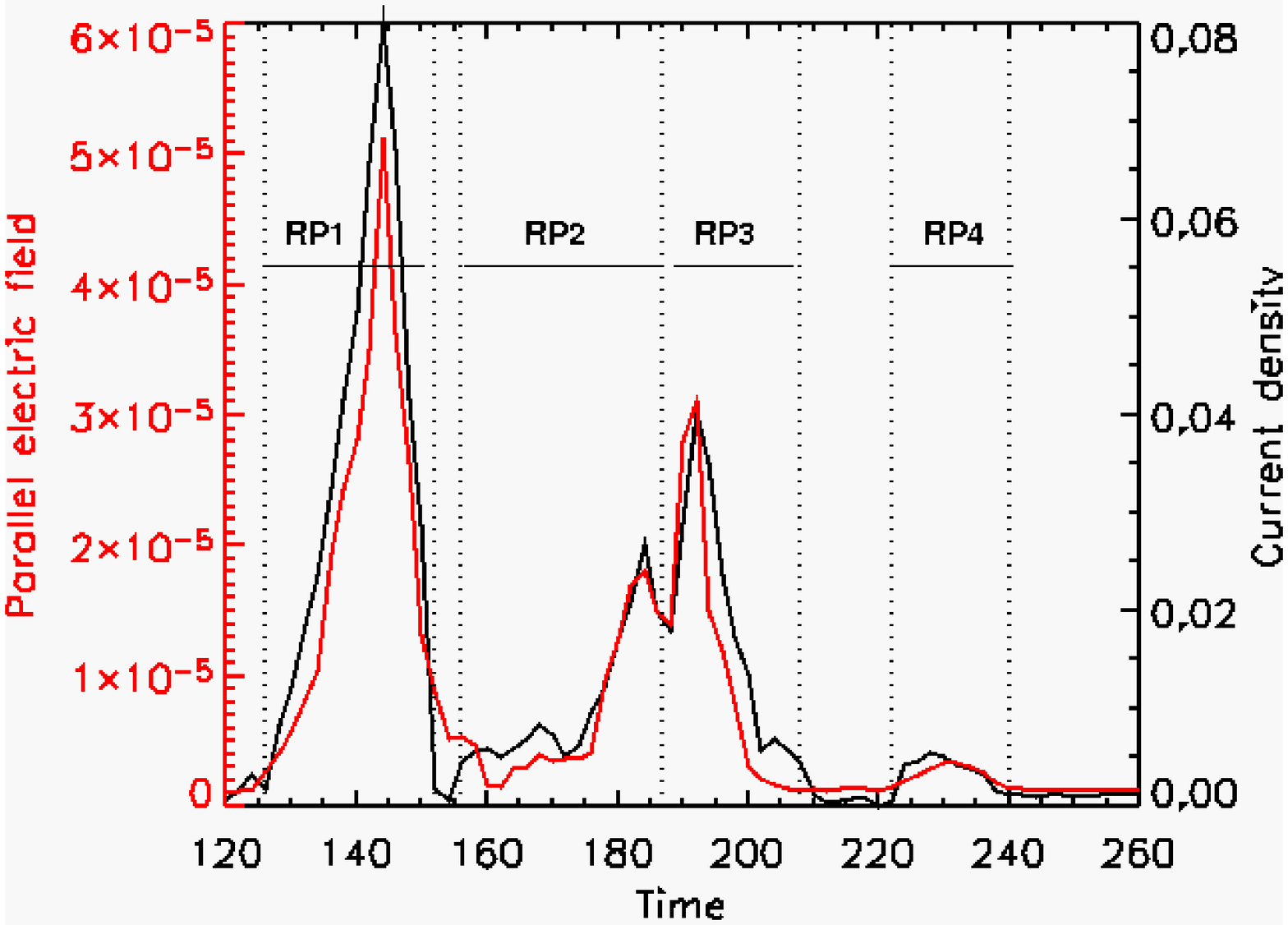}
\includegraphics[width=7.8cm,height=6.cm]{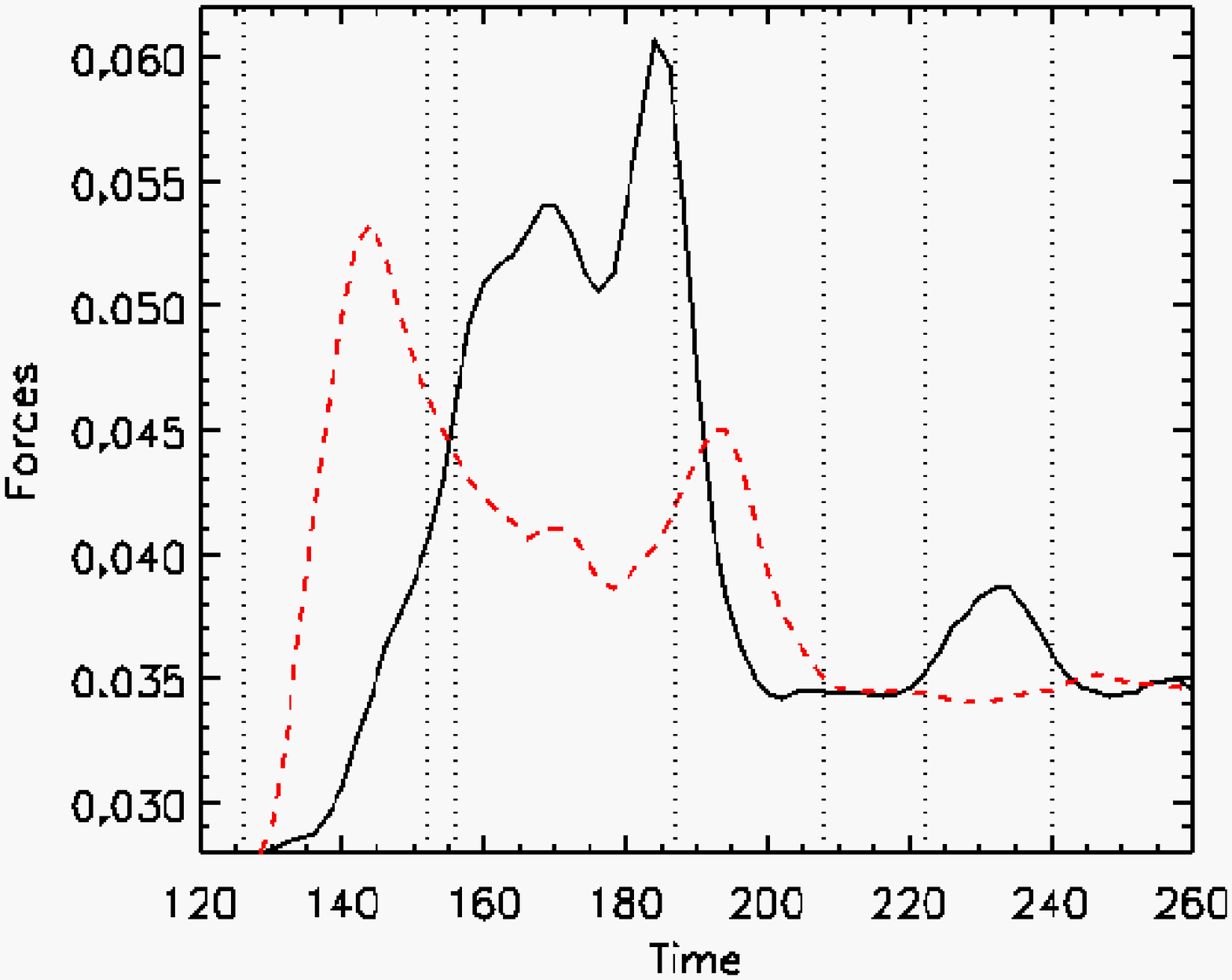}
\end{center}
\caption{{\bf Top}: Time evolution of the maximum {\bf J} (black line) and $E_{\parallel}$ (red line).{\bf Bottom}: Time evolution of the
Lorentz force (dashed) and the total pressure gradient (solid).}
\label{fig3.fig}
\end{figure}
To study the mechanism that drives the recurrency of jets we investigated the forces around the diffusion region.
More precisely, we calculated the maximum values of the Lorentz force and the total (magnetic and gas) pressure gradient within a
3d sub-volume of the {\it arcade} field, in a
very close vicinity of the current structure (e.g., $6<x<14, -5<y<-9, 22<z<24$ at $t=144$).
At the initial stage of RP1, the Lorentz force in the {\it arcade} increases. This is mainly due to
the tension force of the bent magnetic fieldlines, which are accumulated on the {\it arcade} during reconnection.
The tension force is directed towards the outflow region and, thus acts against a possible upward motion of the
{\it arcade} field. The stretching of the fieldlines in the {\it arcade} is apparent in Fig. 1 (panel 1b).
Reconnected fieldlines are added to both outflow regions ({\it arcade} and {\it envelope}), increasing
also the compression and the total pressure there. Thus, the total pressure gradient
increases, although with a slower rate compared to the Lorentz force. The pressure gradient is directed towards the inflow regions and their
interface. At the later stage of RP1, the pressure
gradient continues to increase and eventually becomes larger than the Lorentz force. This change in the forces signals the
onset of the initial stage of RP2. During this phase, the pressure gradient overwhelms the Lorentz force and causes the {\it envelope}
and {\it arcade} fields to reconnect. The inflow regions in RP1, were pulled apart and reconnection between them has been stopped.
After $t=184$, the current structure diffuses away and the pressure gradient in the
inflow (outflow) regions decreases (increases). Consequently, the {\it arcade} field retreats. It is actually shoved aside by the toroidal loops,
which have regained enough stress to push against each other for the second time. The re-joining of the loops occurs: firstly, because
the outward acting pressure gradient force increases during RP2 and secondly, because the magnetic field
of the {\it following} loop continues to emerge and expand laterally.\\
A similar evolution of the forces occurs in the last two reconnection phases. This suggests that the same mechanism underlies
the successive reconnection events and the recurrency of jets. It is the work of the total pressure force against the work of
the Lorentz force on the four magnetic domains, which is responsible for the persistent behavior of the system.
The difference between the early and late reconnection phases
is that the amplitude of the forces is smaller.
This can be understood as follows: it has been shown \citep{arc08b} that the expansion of the emerging field cannot continue for ever.
Eventually, the amount of emerging flux is exhausted and the dynamical rise slows down and reaches an equilibrium. In the present experiments, 
this equilibrium occured for the leading tube at $t\approx 100$.
The emergence of the {\it following} loop and the contact with the pre-existing field is an event that causes an initial disturbance of this equilibrium.
The disturbance leads successive reconnection events, which occur with less efficiency over time, and the overall system approaches 
a new equilibrium stage.

\section {Conclusion}
\label{sec:disc}
Although our results are based on a single experiment as a first approach, it is clear in identifying the important processes and effects
and in establishing the connection between flux emergence, reconnection and recurrent jets in a 3D environment. We expect observations to test whether
recurrent jets appear in active regions due to successive reconnection events triggered by flux emergence and whether such magnetic systems
evolve towards equilibrium. In a forthcoming study, we will investigate the recurrency of jets in a broader range of interacting systems.

\begin{acknowledgements}
Financial support by the European Comission through the SOLAIRE network
(MTRM-CT-2006-035484) is gratefully acknowledged. UKMHD consortium cluster funded by STFC and a SRF grant to the University of St Andrews.
\end{acknowledgements}

\bibliographystyle{aa}

\end{document}